\documentclass{article}
\usepackage{amssymb}

\usepackage{graphicx}
\usepackage{amsmath}

\input{tcilatex}

\begin{document}

\title{Mean Field Theory of Spherical Gravitating Systems}
\author{Peter J. Klinko and Bruce N. Miller \\
Department of Physics, Texas Christian University, \\
Fort Worth, Texas 76129}
\date{\today}
\maketitle

\begin{abstract}
Important gaps remain in our understanding of the thermodynamics and
statistical physics of self-gravitating systems. Using mean field theory,
here we investigate the equilibrium properties of several spherically
symmetric model systems confined in a finite domain consisting of either
point masses, or rotating mass shells of different dimension. We establish a
direct connection between the spherically symmetric equilibrium states of a
self-gravitating point mass system and a shell model of dimension 3. We
construct the equilibrium density functions by maximizing the entropy
subject to the usual constraints of normalization and energy, but we also
take into account the constraint on the sum of the squares of the individual
angular momenta, which is also an integral of motion for these symmetric
systems. Two new statistical ensembles are introduced which incorporate the
additional constraint. They are used to investigate the possible occurrence
of a phase transition as the defining parameters for each ensemble are
altered.
\end{abstract}

\section{Introduction}

The observation that a number of different types of astronomical objects
appear to be in thermodynamically relaxed states has motivated theorists to
understand the thermodynamics and statistical physics of self-gravitating
systems. Of particular note are the globular clusters, consisting of about a
million stars. Besides having relaxed cores, these structures appear to be
organized in two distinct classes characterized by radically different
density profiles, x-ray production, and other features.\cite{Heggie} This
suggests that globular clusters may exist in different thermodynamic phases.
However, in contrast with normal ''chemical'' systems, which have been
successfully described by thermodynamics at the macroscopic level, both the
infinite range and short distance singularity of the Newtonian gravitational
potential introduce problems in the statistical theory of phase transitions
which make their analysis a challenging task. The description of the system
can be simplified by going to the Vlasov limit, i.e. by letting the number
of particles become large while controlling the total mass, M, and energy,
E. In this limit the system is described by the single particle density $%
f(x,v,t)$ in the $\mu $ (position, velocity) space, which is employed by
most of the standard treatments, including the present work.\cite{binney} We
refer to this reduced description as mean field theory (MFT). While MFT
avoids the problem of dealing with an N body formulation, the difficulties
introduced by the singularity and long range of the potential persist.

In the early sixties Antonov investigated the equilibrium behavior of
isolated gravitational systems in MFT.\cite{Antonov} To circumvent the
problem of escape, he confined the mass to a finite region by introducing a
rigid wall. By fixing the total mass and energy, he showed that maximum
entropy solutions for $f$ are spherically symmetric in position and have the
expected Maxwellian velocity dependence. However, he proved that there is no 
\textit{global} maximum to the entropy: while extremal solutions can exist,
these are at best \textit{local} maxima. Antonov \cite{Antonov}, as well as
Lynden-Bell and Wood \cite{lynden1}, closely investigated the spherically
symmetric system confined in a sphere and showed that when the mass and
energy are fixed there are no entropy extrema above a critical radius ($%
R\gtrsim -0.335GM^{2}/E$) . When the radius is less than this value, the
stability of the extremum solutions were studied by several authors
(Lynden-Bell and Katz \cite{lynden2}, Katz \cite{katz1} \cite{katz2},
Padmanabhan \cite{padman} \cite{padman2}, and Bavaud \cite{Bavaud}). They
found that, in general, above a critical density contrast ($\rho \left(
0\right) /\rho \left( R\right) \gtrsim 709$) all extrema are unstable, i.e.
they are not \textit{local} maxima. Lynden-Bell and Wood coined this
phenomena the \textit{gravothermal catastrophe }and it is also referred to
as the Antonov instability. In such a system, there is no upper bound on the
entropy and $\ $a state of arbitrarily large entropy can be constructed from
a centrally concentrated density profile by shifting more of the mass
towards the center (core-halo structures have higher entropy).

More recently, Kiessling has investigated the thermodynamic stability of the
full N body point mass system confined in a spherical box using the
canonical ensemble. To avoid the short range singularity he regularized the
Newtonian interaction by softening it (letting the potential approach a
finite value as the origin is closely approached). He showed that in the
limit that the softening vanishes (Newtonian interaction), the canonical
equilibrium measure is the superposition of Dirac measures at any
temperature, meaning that the system is in a collapsed point mass state. We
have to emphasize that this is the equilibrium solution when the system is
in thermal equilibrium with a heat bath. Also, based on the results of the
finite N-particle system, with proper scaling of the particle mass as we
take the mean-field limit, he showed that the single-particle density
function is proportional to the Dirac distribution. Therefore the system's
equilibrium state is the collapsed state in the canonical ensemble.
Kiessling's conclusions do not contradict the earlier work applying MFT to
the microcanonical ensemble (fixed mass and energy) described above, since
no global entropy maximum was found in that case either.

In fact, the pure Newtonian potential is never a correct picture in general
because of the finite size of stars and atoms. In a nearly hidden appendix
of their paper describing the gravothermal catastrophe, it was first pointed
out by Lynden-Bell and Wood that if we modify the singular $1/r$ Newtonian
gravitational potential at the center by introducing a small minimal
distance between the particles (the so called hard-sphere model) that
complete collapse would be avoided and a global entropy maximum should
exist. \cite{lynden1} They further conjectured that in this situation a
first order phase transition to a centrally concentrated core-halo
configuration would occur as the system energy is reduced. Several authors
demonstrated the existence of a phase-transition in special mean-field
models with a modified gravitational potential. Hertel and Thirring \cite
{hertel} were the first to show \thinspace analytically that a gravitating
system can undergo a first-order phase transition. Although their system is
not purely gravitational and has no singularity, the pair-interaction
potential is purely attractive and has a fixed value when the pair of
particles are in a given sub-domain. Also Lynden-Bell and Lynden-Bell\cite
{lynden3} showed the occurrence of a first-order phase transition in their
special gravitational system, consisting of point particles distributed on a
shell which cannot shrink into a point mass (inner boundary) or expand to
infinity (outer boundary). Kiessling et. al. also applied the hard-sphere
model \cite{kiessling2} to study planet formation: They were able to explain
the existence of observable planets by showing that the mass belonging to
the condensed phase is well below the Jeans mass.\cite{binney} For finite
N-body systems, a gravitational first-order phase transition was first
observed dynamically by Youngkins and Miller \cite{miller}. They
investigated a model consisting of irrotational, concentric, spherical mass
shells confined between two rigid spherical boundaries. The system was
studied both theoretically in the mean-field limit and numerically by N-body
simulations in the microcanonical, canonical, and grand canonical ensembles.
The analysis for this one dimensional system showed that the system
undergoes a first-order phase transition instead of a gravothermal
catastrophe. As expected in gravitational systems, there were some
discrepancies in the results for different ensembles; however the numerical
N-shell simulations were always in good agreement with the corresponding
mean-field predictions.

Much earlier, Eddington \cite{Eddington} determined the form of the general
stationary solution of the Vlasov equation for a spherical system with an
anisotropic velocity distribution which obeys the Schwarzschild law \cite
{binney}. This model explicitly depends on the square of the angular momenta 
$l^{2}$, but it also includes the isothermal part: $f\left( \mathbf{r},%
\mathbf{v}\right) \varpropto e^{-\beta \epsilon }e^{-\gamma l^{2}}$. The
model was presented much earlier, but it may have been forgotten. Later,
several phenomenological models were improved by including Eddington's
anisotropic term (King-Michie models and others \cite{gunn} \cite{Heggie} 
\cite{binney}), giving a better fit to the observed density profiles of
globular clusters. However, in some cases, a good fit was not obtained for
globular clusters with core-halo structures. In other words, a fair number
of globular clusters which have a small and very dense core surrounded by a
thin halo structure, do not obey these empirical density-fit models. On the
other hand, in addition to the system energy, the sum of the angular momenta
squares $L_{2}=\sum l_{i}^{2}$ is also an integral of the motion for any
isolated spherically symmetric system in the mean-field limit \cite{binney},
and should not be ignored.

The purpose of our work is to present a more general approach for
investigating the equilibrium properties of confined, spherical, systems
then those of Antonov, Lynden-Bell and Wood, etc. mentioned above which
takes into account both integrals of a spherical gravitating system, and to
introduce idealized dynamical ''shell'' models which also satisfy these
constraints. Model gravitating systems consisting of a collection of
concentric, infinitesimally thin, spherical shells were first introduced by
Henon \cite{Henon}. They are useful for investigating the initial stages of
evolution of a spherically symmetric self gravitating system, before the
onset of binary formation arising from three body effects. \cite{Heggie}
They have the further advantages of ease and accuracy of algorithm
construction, since it is possible to analytically solve for the motion of
each shell between encounters, eliminating the need for the tedious and
slow, step-wise integration of coupled, nonlinear, differential equations. 
\cite{russian} \cite{miller} \cite{miller2}.

In the present work we consider the mean field theory of a system of
gravitating point particles moving in three dimensional space, as well as
that of thin, rotating, spherical mass shells with angular momentum vectors
restricted to manifolds of one, two and three dimensions. We first determine
conditions for the equilibrium one particle probability density function $%
f\left( x,p\right) $of a unit mass particle (shell) by finding the entropy
extrema with respect to the constraints of (1) the normalization, (2) system
energy $E$, and (3) the sum of the squares of the angular momentum, $L_{2}$.
We then show that the introduction of the new integral $L_{2}$ suggests a
new type of canonical ensemble ($T-\gamma )$, in addition to the extension
of the microcanonical ensemble ($E-L_{2})$ . A nonlinear differential
equation governing the radial density valid for each ensemble is derived for
the case of the three dimensional point mass system, and each shell system.
We then prove that the radial density of the shell system with angular
momentum confined to the Euclidean plane satisfies the identical
differential equation as the three dimensional point mass system, and we
carefully study the equilibrium solutions for this case numerically. The
stability of the extremum solutions of each model is investigated in both
microcanonical ($E-L_{2})$ and canonical ($T-\gamma )$ ensembles. At first
glance it is natural to anticipate that the centrifugal barrier associated
with the additional constraint will eliminate any tendency for complete core
collapse without introducing an inner boundary in the system or changing the
gravitational potential by other means. We conclude by investigating the
possible presence of a phase-transition which would remove the gravothermal
catastrophe .

\section{The Entropy Extrema}

\subsection{Spherically symmetric point mass systems}

The primary goal is to evaluate the equilibrium single particle probability
density function $f\left( x,p\right) $ which maximizes the entropy in the
mean-field limit. Consider the spherically symmetric isolated point mass
system in three space dimensions with total mass $M$ confined in a sphere of
radius $b$. We choose units where $G=M=b=1$ and introduce spherical
coordinates $x=\left( r,\varphi ,\vartheta \right) $ . The Lagrangian per
unit mass of a single particle moving in the mean field potential $\Phi (r)$
(\ref{pot}) is 
\begin{equation}
L=\frac{1}{2}\dot{r}^{2}+\frac{1}{2}r^{2}\left( \dot{\vartheta}^{2}+\dot{%
\varphi}^{2}\sin ^{2}\vartheta \right) -\Phi \left( r\right) ,
\label{lagrange}
\end{equation}
where, for a Newtonian pair-wise interaction, $\Phi \left( r\right) $ is
given by :

\begin{equation}
\Phi \left( r\right) =-\int \int \left[ G\left( r,r^{/}\right) +G\left(
r^{/},r\right) \right] f\left( x^{/},p^{/}\right) d^{3}x^{/}d^{3}p^{/},
\label{pot}
\end{equation}
where 
\begin{equation*}
G\left( r,r^{/}\right) =\frac{\Theta \left( r-r^{/}\right) }{r}
\end{equation*}
and, as usual, $\Theta \left( r\right) $ denotes the Heavyside step
function. The Hamiltonian is then 
\begin{equation}
H=\frac{1}{2}v^{2}+\frac{l_{\vartheta }^{2}}{2r^{2}}+\frac{l_{\varphi }^{2}}{%
2r^{2}\sin ^{2}\vartheta }+\Phi \left( r\right)  \label{hamilton}
\end{equation}
where $p=\left( v,l_{\varphi },l_{\vartheta }\right) $ are the corresponding
canonical momenta. Our plan is first to construct the entropy extrema, and
then verify whether or not the solutions are local maxima. The entropy of
the system is\cite{binney}\cite{padman} 
\begin{equation}
S=-\int \int f\ln fd^{3}xd^{3}p,  \label{pointentropy}
\end{equation}
and the constraints for which we need to find the extremum are: (1)
normalization of $f$, (2) energy conservation in the complete system, and
(3) conservation of $L_{2}$ : 
\begin{equation}
1=\int \int fd^{3}xd^{3}p  \label{norm}
\end{equation}
\begin{equation}
L_{2}=\int \int l^{2}fd^{3}xd^{3}p  \label{L2}
\end{equation}
\begin{equation}
E=\int \int f\left( \frac{1}{2}v^{2}+\frac{1}{2r^{2}}l^{2}+\frac{1}{2}\Phi
\left( r\right) \right) d^{3}xd^{3}p  \label{E}
\end{equation}
where

\begin{equation*}
l^{2}=\left( l_{\vartheta }^{2}+\frac{l_{\varphi }^{2}}{\sin ^{2}\vartheta }%
\right) .
\end{equation*}
Introducing Lagrange multipliers $\alpha ,$ $\beta ,$ $\gamma $ we have an
extremum for the functional $S$ when 
\begin{equation}
0=\delta \left( S-\alpha -\beta E-\gamma L_{2}\right) .  \label{extremum}
\end{equation}
Taking the first variation of each term in (\ref{extremum}), and asserting (%
\ref{norm}), (\ref{E}), (\ref{L2}), and (\ref{pot}), we obtain: 
\begin{equation*}
0=-\int \int \left\{ \left( \ln f+1\right) +\alpha +\beta \left[ \frac{1}{2}%
v^{2}+\frac{1}{2r^{2}}l^{2}+\Phi \left( r\right) \right] +\gamma
l^{2}\right\} \delta fd^{3}xd^{3}p
\end{equation*}
from which finally 
\begin{equation}
0=-\left( \ln f+1\right) -\alpha -\beta \left( \frac{1}{2}v^{2}+\Phi \left(
r\right) \right) -\left( \frac{\beta }{2r^{2}}+\gamma \right) l^{2}.
\end{equation}
Thus the one-particle probability density function (pdf) is 
\begin{equation}
f=\exp [-\left( \alpha +1\right) -\beta \left( \frac{1}{2}v^{2}+\Phi \right)
-\left( \frac{\beta }{2r^{2}}+\gamma \right) l^{2}]..  \label{3dpointdensity}
\end{equation}
In order to obtain the radial density $\rho \left( r\right) $, we have to
integrate $f$ over the other variables. To ensure that the integrals over $v$
, $l_{\vartheta }$, and $l_{\varphi }$ converge, the following conditions
are necessary:$\;\beta >0$ and $\beta /2r^{2}+\gamma >0$ at any $r$.
Therefore the second condition is $2\gamma /\beta >-1/b^{2}=-1$. Using $%
K=\exp \left[ -\left( \alpha +1\right) \right] $ , we get 
\begin{equation*}
\rho \left( r\right) =\int \int fd^{3}pd\varphi d\vartheta =\int_{0}^{2\pi
}\int_{0}^{\pi }K\sqrt{\frac{2\pi }{\beta }}\frac{\pi \sin \vartheta }{\frac{%
\beta }{2r^{2}}+\gamma }\exp \left( -\beta \Phi \right) d\varphi d\vartheta
\end{equation*}
\begin{equation}
=\frac{K\left( 2\pi \right) ^{\frac{5}{2}}}{\sqrt{\beta }}\left( \frac{\beta 
}{2r^{2}}+\gamma \right) ^{-1}e^{-\beta \Phi \left( r\right) }
\label{3dpointmassdens}
\end{equation}
and, with the Poisson equation in a spherical coordinate system, 
\begin{equation*}
\Delta \Phi =\frac{1}{r^{2}}\frac{d}{dr}\left( r^{2}\frac{d\Phi }{dr}\right)
=4\pi \rho _{V}
\end{equation*}
where $\rho _{_{V}}$ is the volumetric mass density. In some cases, it's
more convenient to use the linear (radial) density instead of the volume
density: 
\begin{equation}
\frac{d}{dr}\left( r^{2}\frac{d\Phi }{dr}\right) =\rho \left( r\right) .
\label{poisson}
\end{equation}
Note that because $M=1$, the radial probability density function and the
linear mass density function are the same. Introducing a new function $\Psi
=\beta \Phi $ and employing \ref{3dpointmassdens} we can rewrite (\ref
{poisson}) as 
\begin{equation*}
\frac{d}{dr}\left( r^{2}\frac{d\Psi }{dr}\right) =\rho \left( r\right)
=K\left( 2\pi \right) ^{\frac{5}{2}}\sqrt{\beta }\left( \frac{\beta }{2r^{2}}%
+\gamma \right) ^{-1}e^{-\Psi \left( r\right) }
\end{equation*}
obtaining a closed equation for $\Psi .$ This in turn can be simplified by
introducing constants $C$ and $\Gamma ,$ 
\begin{equation}
\frac{d}{dr}\left( r^{2}\frac{d\Psi }{dr}\right) =C\left( \frac{1}{r^{2}}%
+\Gamma \right) ^{-1}e^{-\Psi \left( r\right) },  \label{3dpointfinal}
\end{equation}
\begin{equation}
C=K\left( 2\pi \right) ^{\frac{5}{2}}\frac{2}{\sqrt{\beta }}>0,
\end{equation}
\begin{equation*}
\Gamma =\frac{2\gamma }{\beta }>-1.
\end{equation*}
This is the final form of the differential equation for the scaled potential
which we will solve using numerical methods. Finally, we can evaluate $E$, $%
L_{2}$, and $S$ in terms of the Lagrange multipliers, the density $\rho
\left( r\right) $, and the potential $\Phi \left( r\right) .$ After
integrating (\ref{E}), (\ref{L2}), and (\ref{pointentropy}), we obtain: 
\begin{equation}
S=\alpha +\frac{5}{2}+\beta \int_{0}^{b}\rho \Phi dr  \label{3dpointentropy}
\end{equation}
\begin{equation}
L_{2}=\int_{0}^{b}\rho \left( \frac{\beta }{2r^{2}}+\gamma \right) ^{-1}dr
\label{3dpointL2}
\end{equation}
\begin{equation}
E=\frac{1}{2\beta }+\int_{0}^{b}\rho \left( \frac{1}{\beta +2\gamma r^{2}}+%
\frac{\Phi }{2}\right) dr  \label{3dpointE}
\end{equation}

\subsection{Shell systems}

In this section we consider the system of spherically symmetric,
infinitesimally thin, mass shells confined in a sphere with radius $b.$ We
can define three basic types of the model with dimensions $d=2,3,4$. The
one-dimensional $d=1$ non-rotating shell system is discussed in \cite{miller}
. In the $d=2$ case, every shell rotates about a fixed axis. When $d=3,$ the
rotational axis of every shell is in a fixed plane, while $d=4$ means that
every shell can rotate about any arbitrary axis. Therefore we use $d-1$
angles as coordinates. Let us consider these systems in the mean-field limit
again using units where $M=G=b=1$. With the potential discussed above (\ref
{pot}), the Lagrangian and the Hamiltonian of a unit mass shell in a
spherical coordinate system are, respectively,: 
\begin{equation}
L=\frac{1}{2}\overset{.}{r}^{2}+\frac{1}{3}r^{2}\sum_{k=1}^{n}\dot{\varphi}%
_{k}^{2}-\Phi \left( r\right)
\end{equation}
\begin{equation}
H=\frac{1}{2}v^{2}+\frac{3}{4}\frac{\sum_{k=1}^{n}l_{k}^{2}}{r^{2}}+\Phi
\left( r\right)
\end{equation}
where $\varphi _{k}$ $\left( k=1,2,3\right) $ are the angles around $x,$ $y$
and $z,$ the $l_{k}$ are the $x,y,z$ components of the angular momentum per
unit mass, and we have used the fact that the moment of inertia of a shell
with unit mass is $2/3r^{2}.$ In the equations above, we use $n=d-1$ which
is the number of degrees of freedom coming only from rotation. This is a
generalization of the previous model, and so is the method to find the
equilibrium pdf, $f\left( x,p\right) $. As in the previous section, in order
to get the equilibrium solutions first we have to find the extremum of the
entropy: 
\begin{equation}
S=-\int \int f\ln fd^{n+1}xd^{n+1}p  \label{nshellS}
\end{equation}
with respect to the three constraints of normalization, $E,$ and $L_{2}:$%
\begin{equation*}
1=\int \int fd^{n+1}xd^{n+1}p,
\end{equation*}
\begin{equation}
L_{2}=\int \int f\left( \sum_{k=1}^{n}l_{k}^{2}\right) d^{n+1}xd^{n+1}p,
\label{nshellL2}
\end{equation}
\begin{equation}
E=\int \int f\left( \frac{1}{2}v^{2}+\frac{3}{4}\frac{\sum_{k=1}^{n}l_{k}^{2}%
}{r^{2}}+\frac{1}{2}\Phi \left( r\right) \right) d^{n+1}xd^{n+1}p.
\label{nshellE}
\end{equation}
The solution for the variational problem can be obtained easily, and the
one-particle density function is now 
\begin{equation}
f=\exp \left[ -\left( \alpha +1\right) \right] \exp \left[ -\beta \left( 
\frac{1}{2}v^{2}+\Phi \right) \right] \exp \left[ -\left( \frac{3\beta }{%
4r^{2}}+\gamma \right) \sum_{k=1}^{n}l_{k}^{2}\right] .
\label{nshelldensity}
\end{equation}
Therefore the radial mass density function is 
\begin{equation}
\rho \left( r\right) =\int \int fd^{n+1}pd^{n}\varphi =K\left( 2\pi \right)
^{\frac{2n+1}{2}}\frac{\pi ^{\frac{n}{2}}}{\sqrt{\beta }}\left( \frac{3\beta 
}{4r^{2}}+\gamma \right) ^{-\frac{n}{2}}\exp \left( -\beta \Phi \right) .
\end{equation}
From the Poisson equation, again using using $\Psi =\beta \Phi ,$ we find, 
\begin{equation}
\frac{d}{dr}\left( r^{2}\frac{d\Psi }{dr}\right) =K\left( 2\pi \right) ^{%
\frac{2n+1}{2}}\pi ^{\frac{n}{2}}\sqrt{\beta }\left( \frac{3\beta }{4r^{2}}%
+\gamma \right) ^{-\frac{n}{2}}\exp \left( -\Psi \right)
\end{equation}
which can be simplified to 
\begin{equation}
\frac{d}{dr}\left( r^{2}\frac{d\Psi }{dr}\right) =C\left( \frac{1}{r^{2}}%
+\Gamma \right) ^{-\frac{n}{2}}e^{-\Psi \left( r\right) },
\label{nshellfinal}
\end{equation}
\begin{equation}
C=\frac{2^{\frac{4n+1}{2}}}{(3\beta )^{\frac{n}{2}}}\pi ^{\frac{3n+1}{2}}K%
\sqrt{\beta }>0,
\end{equation}
\begin{equation}
\Gamma =\frac{4\gamma }{3\beta }>-1.
\end{equation}

Comparing the results (\ref{nshellfinal}) for the 3 dimensional shell system
($n=2$) to that of a 3 dimensional point mass system (\ref{3dpointfinal}),
we can see that these two systems are equivalent in so far as (\ref
{3dpointfinal} ) and (\ref{nshellfinal}) have the same form.

It is useful to evaluate $S,L_{2},$ and $E$ in terms of the Lagrange
multipliers, $\rho \left( r\right) $, and $\Phi \left( r\right) .$ The
integration of (\ref{nshellS}), (\ref{nshellL2}), and (\ref{nshellE}) yields 
\begin{equation}
S=\alpha +\frac{n+3}{2}+\beta \int_{0}^{b}\rho \Phi dr
\end{equation}
\begin{equation}
L_{2}=\frac{n}{2}\int_{0}^{b}\rho \left( \frac{3\beta }{4r^{2}}+\gamma
\right) ^{-1}dr
\end{equation}
\begin{equation}
E=\frac{1}{2\beta }+\int_{0}^{b}\rho \left( \frac{3n}{8}\frac{1}{\frac{3}{4}%
\beta +\gamma r^{2}}+\frac{\Phi }{2}\right) dr.
\end{equation}

\section{T-$\protect\gamma $ ensemble}

In the previous sections we derived expressions for the entropy extremum
solutions for our model systems in terms of the local radial density. A
third Lagrange multiplier,$\gamma $, was introduced to satisfy the new
constraint on $L_{2}$ $.$ Thus specifying $E$ and $L_{2}$ (and, of course, $%
M $) defines the analogue to the microcanonical ensemble for these systems.
Alternatively, by fixing $\beta $ and $\gamma $, we can define the analogue
of the canonical ensemble, in which $E$ and $L_{2}$ are not fixed, but their
average is determined by $\beta $ and $\gamma $ , where $\gamma =\frac{%
\partial S}{\partial L_{2}}$. We call this the $T-\gamma $ ensemble. It can
be modeled by imagining that the system is in contact with a heat bath with
constant $\beta $ and also an $l^{2}$ bath at constant $\gamma $. The new $%
l^{2}$ bath corresponds to the fact that we allow some $l^{2}$ exchange
between the system and the bath. Since the system is spherically symmetric
in position, while the system can also exchange angular momentum with the
bath, its vector average will vanish. As an example, we can imagine a
globular cluster which is embedded in some large spherically symmetric
stellar neighborhood with an isotropic velocity distribution. The mean
angular momentum of both system and bath is zero, and only $l^{2}$ and
energy exchange can occur (for the moment, we do not take into account the
possibility that particles can escape from the cluster).

In performing calculations it is more convenient to use the $T-\gamma $
ensemble than the $E-L_{2}$ microcanonical ensemble because in the latter we
have to find the Lagrange multipliers from the given $E$ and $L_{2}$. As can
be seen from \ref{3dpointentropy}, \ref{3dpointL2}, and \ref{3dpointE} this
is a nontrivial and laborious task. In this ensemble the relevant
thermodynamic potential is an extension of the Helmholz free energy: 
\begin{equation}
F=E-\frac{1}{\beta }S+\frac{\gamma }{\beta }L_{2}
\end{equation}
and equilibrium states, if they exist, minimize $F.$

\section{Stability}

From the variational problem, we only know the extremum solutions. In order
to separate the unstable solutions from the locally stable, we use the
modified method of Poincare's linear series of equilibria. Following Katz, 
\cite{katz1}\cite{katz2} we can generalize the method to the case of a
functional. From now on, we discuss the generalized version of the method
which has to be applied to our models to determine whether an extremum
solution is stable or unstable. We outline the method below. Let's assume
that we want to find the maximum of the functional $F^{*}\left( f,s\right) $
which depends parametrically on $s.$ The function $f:\Omega \rightarrow R$,
where $\Omega $ is a compact domain $\Omega \in R^{3}$, and the local
maximum of the functional is the stable solution. Suppose we partition $%
\Omega $ and consider the vector $x\in R^{n}$ with elements $x_{i}=f\left(
y_{i}\right) $ ($y_{i}\in \Omega $ is not a coordinate but an indexed
element in $\Omega $). Instead of dealing with the functional $F^{*}$, we
can construct a function $F:R^{n}\rightarrow R$ such that $F\left(
x,s\right) \approx F^{*}\left( f,s\right) $. The accuracy can be controlled
by refining the partition (increasing n). The problem is then shifted to
finding the extrema of $F$: 
\begin{equation}
\partial _{i}F\left( x,s\right) =0
\end{equation}
Denote the extremum solutions of the problem by $x=\left\{ X_{a}\left(
s\right) \right\} $ where $a,b=1...N$, labels different extremal solutions.
Assume that the first and second derivatives of $F$ are continuous in $x$
and $s,$ and $\dot{X}_{a}$ is continuous as well. Assume also that the
matrix $\left( -\partial _{i}\partial _{j}F\right) _{a}$ has a
non-degenerate eigenvalue spectrum, which we may consider to be ordered: $%
k_{1a}\left( s\right) <k_{2a}\left( s\right) <$ . . .$<k_{na}\left( s\right)
<$ ..., and further that $\left( -\partial _{i}\partial _{j}F\right) _{a}$
is diagonal. (We can always transform it into that form.). Let's evaluate $F$
at the extremum points $x=X_{a}\left( s\right) $ where 
\begin{equation}
\partial _{i}F\left( X_{a},s\right) =0.
\end{equation}
Then, as the parameter $s$ is varied, on the extremum labeled $a$, 
\begin{equation}
0=\frac{d}{ds}\partial _{i}F\left( X_{a},s\right) =\left( \partial
_{s}\partial _{i}F\right) _{a}+\sum_{j=1}^{\infty }\left( \partial
_{j}\partial _{i}F\right) _{a}\dot{X}_{a}^{j}=\left( \partial _{s}\partial
_{i}F\right) _{a}-k_{ia}\left( s\right) \dot{X}_{a}^{i},
\end{equation}
\begin{equation}
\dot{X}_{a}^{i}=\frac{\left( \partial _{s}\partial _{i}F\right) _{a}}{
k_{ia}\left( s\right) }.
\end{equation}
The stability of the extremum is determined by the second derivative of $%
F_{a}\left( s\right) =F\left( X_{a},s\right) $, 
\begin{equation}
\ddot{F}_{a}=\left( \partial _{s}^{2}F\right) _{a}+\sum_{i=1}^{\infty
}\left( \partial _{i}\partial _{s}F\right) _{a}\dot{X}_{a}^{i}=\left(
\partial _{s}^{2}F\right) _{a}+\sum_{i=1}^{\infty }\frac{\left( \partial
_{s}\partial _{i}F\right) _{a}^{2}}{k_{ia}}.  \label{stability}
\end{equation}
The stability will change only when one of the $k_{ia}$ changes sign, and a
change in stability occurs only at bifurcation or limit points \cite{katz1} 
\cite{katz2}. Therefore we have to investigate the dependence of $\ddot{F}%
_{a}$ on $s$ in order to decide how the stability changes at an $a-b$
bifurcation or limit point . If we look at (\ref{stability}), one can see
that when a particular $k_{ia}\left( s_{0}\right) $ changes sign from
positive to negative, at that point $\underset{s\rightarrow s_{0}}{\lim }%
\ddot{F}_{a}=+\infty $ and similarly, from the other $b$ branch, $\underset{%
s\rightarrow s_{0}}{\lim }\ddot{F}_{b}=-\infty $.

Going back to our original problem, we have to apply the method to our model
systems both in the $E-L_{2}$ and $T-\gamma $ ensembles. The only difficulty
is that the method discussed above only allows us to include systems with
one control parameter, $s$. In our case, we are free to fix either $E$ or $%
L_{2}$ in the $E-L_{2}$ ensemble and $\beta $ or $\gamma $ in the $T-\gamma $
ensemble and regard the other as the control parameter. Afterwards, we can
change the previously fixed parameter and apply the method again for a wide
range of parameter sets. In the $E-L_{2}$ ensemble a natural choice of
parameter is $E$ with one fixed value of $L_{2}$, as the entropy has to be a
local maxima if the system is in a locally stable state. If we are at the
extremum solution points 
\begin{equation}
\dot{S}=\frac{dS}{dE}=\frac{\partial S}{\partial E}=\beta .
\end{equation}
For a different value of $L_{2}$, we can clearly see which branch of the
extremum solutions are unstable. To obtain a complete description we can use
the same method if $E$ is fixed: 
\begin{equation}
\dot{S}=\frac{dS}{dL_{2}}=\frac{\partial S}{\partial L_{2}}=\gamma .
\end{equation}
In the $T-\gamma $ ensemble, first consider the case where we fix $\gamma ,$
and we are looking for the maximum of the functional $-\beta F$ . On a
branch of the extremum solutions 
\begin{equation}
\frac{dF}{d\beta }=\frac{\partial E}{\partial \beta }+\frac{S}{\beta ^{2}}-%
\frac{1}{\beta }\frac{\partial S}{\partial E}\frac{\partial E}{\partial
\beta }-\frac{1}{\beta }\frac{\partial S}{\partial L_{2}}\frac{\partial L_{2}%
}{\partial \beta }-\frac{\gamma }{\beta ^{2}}L_{2}+\frac{\gamma }{\beta }%
\frac{\partial L_{2}}{\partial \beta }.
\end{equation}
Using $\frac{\partial S}{\partial E}=\beta $ and $\frac{\partial S}{\partial
L_{2}}=\gamma $, we easily find 
\begin{equation*}
\frac{dF}{d\beta }=\frac{S}{\beta ^{2}}-\frac{\gamma }{\beta ^{2}}L_{2},
\end{equation*}
\begin{equation}
-\frac{d\left( \beta F\right) }{d\beta }=-\left( F+\beta \frac{dF}{d\beta }%
\right) =-E.
\end{equation}
If we construct the stable branches for several values of $\gamma $, we can
build up a general picture of the stability of the extremum solutions. We
can apply the method for the case of a fixed $\beta $ as well. The extremum
solutions are stable when $-\beta F$ is maximum: 
\begin{equation}
\frac{dF}{d\gamma }=\frac{\partial E}{\partial \gamma }-\frac{1}{\beta }%
\frac{\partial S}{\partial E}\frac{\partial E}{\partial \gamma }-\frac{1}{
\beta }\frac{\partial S}{\partial L_{2}}\frac{\partial L_{2}}{\partial
\gamma }+\frac{1}{\beta }L_{2}+\frac{\gamma }{\beta }\frac{\partial L_{2}}{%
\partial \gamma },
\end{equation}
\begin{equation*}
\frac{dF}{d\gamma }=\frac{1}{\beta }L_{2},
\end{equation*}
\begin{equation}
-\frac{d\left( \beta F\right) }{d\gamma }=-L_{2}.
\end{equation}

From the results derived above we have an easy to apply tool to separate the
unstable solutions. In the microcanonical ($E-L_{2}$) description, we have
to inspect the extremum solutions in the $\beta -E$ plane for several fixed
values of $L_{2}$ or in the $\gamma -L_{2}$ plane for several fixed values
of $E$. However, in the $T-\gamma $ ensemble, we have to inspect the
extremum solutions in the $\left( -E\right) -\beta $ plane for fixed $\gamma 
$ values, or we have to consider the extremum solutions in the $\left(
-L_{2}\right) -\gamma $ plane for the case of fixed $\beta $. Either way, as
we will show below, we can generalize the approach to two parameters.

\section{ Numerical method}

In order to find the entropy extremum solutions for the systems above, we
have to solve (\ref{3dpointfinal}) or (\ref{nshellfinal}), for the 3D point
mass system, or the three types of shell system. Generally, all of the
differential equations can be written in the form of 
\begin{eqnarray}
\frac{dy}{dr} &=&C\left( \frac{1}{r^{2}}+\Gamma \right) ^{-\frac{n}{2}%
}e^{-\Psi \left( r\right) }  \label{genfinal} \\
\frac{d\Psi }{dr} &=&\frac{y}{r^{2}}  \notag
\end{eqnarray}
where $\Psi =\beta \Phi $ and $n=d-1$ (as above, d is the dimension of the
system). Of course, the parameters $\Gamma >-1$ and $C>0$ are different case
by case. But it's quite interesting to mention that these systems behave
similarly, regardless of whether we deal with a point mass system or a shell
system, as long as the dimensions are the same. This is no longer a surprise
if we look back at the Hamiltonians. But the similarity does also mean that,
for example, a three dimensional point mass system is equivalent to a three
dimensional shell system as far as the one particle density function and the
stability of solutions are concerned. Also another advantage of the analogy
is that we can dynamically model a three dimensional point mass system with
a 3 dimensional shell system. Both systems should show the same equilibrium
properties in the mean-field limit. Of course, we have to reassign the
momentum of inertia of a shell to a different value in order to get \textit{%
exactly} the same Hamiltonian in both cases.

Equation (\ref{genfinal}) should be solved with the following boundary
conditions: 
\begin{eqnarray}
y\left( 0\right) &=&0  \label{boundarycond} \\
\Psi \left( 1\right) &=&-\beta  \notag
\end{eqnarray}
Unfortunately, there are two things which make finding the solutions more
difficult. First of all, (\ref{genfinal}) has a singularity at $r=0$ and,
secondly, the existence and uniqueness of the solution are questionable for
any given $C$, $\beta $, and $\Gamma .$ We cannot simply set $C$ and $\Gamma 
$ to satisfy our constraints of specified $E$ and $L_{2}$ because we don't
have explicit forms of $E$ and $L_{2}$ in terms of $C$ and $\Gamma $ : the
constraints are functionals of $\rho $ and $\Phi $. To eliminate this
problem we can change our boundary condition problem to an initial value
problem because, in the latter case, we can ensure the existence and
uniqueness of our solution for any $C>0$ and $\Gamma >-1$. Therefore we
choose 
\begin{eqnarray}
y\left( 0\right) &=&0  \label{initialvalue} \\
\Psi \left( 0\right) &=&\Psi _{0}  \notag
\end{eqnarray}
where $\Psi _{0}\in \left( -\infty ,+\infty \right) $, and we are able to
construct the solution around $r=0$ in the form of a power series of $\Psi $
and $y$. From the numerical point of view, we should take the power series
of $\Psi $ and $y$ in $r_{1}$ (a sufficiently small radius around $r=0$),
and then continue the integration of (\ref{genfinal}) numerically from that
point with the new initial values 
\begin{eqnarray}
y\left( r_{1}\right) &=&y_{01} \\
\Psi \left( r_{1}\right) &=&\Psi _{01}  \notag
\end{eqnarray}
Although the solution we get from (\ref{initialvalue}) is unique, it does
not satisfy both (\ref{boundarycond}) and normalization. However, we can
prove that we can't find any $\chi _{1},\chi _{2}\in V_{0}$ such that $\Psi
_{1}=\Psi _{2}+c$, where $V_{0}$ consists of those solutions $\chi =\left(
y,\Psi \right) $ which satisfy the same initial value problem of (\ref
{genfinal}) and (\ref{initialvalue}) with $C=C_{0}>0$ and any $\Gamma >-1$.
Therefore all of the solutions are unique with fixed $C$ and $\Gamma $ or,
in other words, all $\chi \in V_{0}$ are unique. Now let's consider $V_{1}$,
which can be defined the same way as we defined $V_{0}$, but with $%
C=C_{1}\neq C_{0}$. We can also prove that for any $\chi _{2}\in V_{1}$,
there is only one $\chi _{1}\in V_{0}:\chi _{2}=\chi _{1}-\left( 0,c\right) $%
. Thus we can find all of the physically unique solutions by solving the
initial value problem with a fixed $C_{0}.$ In practice, we used the
Bulirsch-Stoer method \cite{Numrec} to integrate the coupled, nonlinear,
differential equations from $y_{1}$ and Bode's five point method \cite
{Numrec} to evaluate integrals. Relative errors were controlled to within 10$%
^{-12}$.\newpage

\FRAME{ftFU}{3.9902in}{3.0502in}{0pt}{\Qcb{The extremum solution curves at
different values of $\Gamma =2\protect\gamma /\protect\beta $ (from the top
to the bottom $\Gamma =-0.9,-0.5,0,2,5.$) Note that each curve has an upper
bound $\protect\beta _{c}\left( \Gamma \right) ,$ and no extrema can be
found with $\protect\beta >\protect\beta _{c}\left( \Gamma \right) $. Also
note that over a certain range of $\protect\beta $, multiple solutions are
present. However from those multiple solutions, only those are stable which
have the highest value of $\Psi _{0}$ (See Figure 2). The asymptotic cases
are $\Psi _{0}\rightarrow -\infty $ ($\protect\beta \rightarrow \protect\beta
_{0}$ and $\protect\rho _{V}\left( r\right) \varpropto 1/r^{2}$) and $\Psi
_{0}\rightarrow \infty $ which is the high temperature limit.}}{\Qlb{FIG01}}{%
fig01.eps}{}

\section{Numerical results}

The numerical solutions of (\ref{genfinal}) are presented in Figure 1 for
the case of the spherically symmetric point mass system. The numerical
results for the other models are qualitatively very similar, so we will use
the point mass system to demonstrate their general features. As we can see,
the solutions are bounded and, as a comparison, the results for the familiar
isothermal sphere model ($\Gamma =0)$\cite{chandra}\cite{padman} are
presented as well. Examination of Figure \ref{FIG01} clearly demonstrates
the existence of an upper bound, $\beta _{c}\left( \Gamma \right) ,$ which
means that below a critical temperature there's no extremum solution at
fixed $\Gamma $. In fact, this can be proved rigorously from the
differential equations.\cite{Igor} We can also see from the graph that $\dot{%
\beta}_{c}=\frac{d\beta _{c}}{d\Gamma }<0$ and that, when $\beta \rightarrow
\beta _{0}$ $=\underset{\Psi _{0}\rightarrow -\infty }{\lim }\beta \left(
\Psi _{0}\right) $ , the number of solutions goes to infinity.\newpage

\FRAME{ftbpFU}{3.9946in}{3.0519in}{0pt}{\Qcb{The extremum solutions for $%
\Gamma =-0.74.$ For $\protect\beta =2.7$, three extremum solutions are
present, but only the one with the largest $\Gamma $ is stable. For
comparison, also see Figure 8.}}{\Qlb{FIG02}}{fig02.eps}{}
In Figure \ref{FIG02}, we show plots of three solutions for a
given temperature (we don't know so far which of them, if any, are stable),
and in Figure \ref{FIG03} we plot the volume density profiles of these
solutions which have been normalized to the central density. As we can see,
solutions represented by smaller $\Psi _{0}$ are more and more concentrated
at the center. As a comparison, we also give the density profiles of the
well known isothermal sphere ($\Gamma =0$ ). There's some difference in the
shape of the density profiles in the case where $\Gamma \neq 0$ but, in
general, the volume density profile is singular as $\Psi _{0}\rightarrow
-\infty ,$ and has the asymptotic solution $\rho _{V}=\frac{1}{4\pi r^{2}}$
. Note that the linear density is $\rho =1$ in this asymptotic case.\newpage

\FRAME{ftbpFU}{3.9946in}{3.0519in}{0pt}{\Qcb{The relative volume density
profiles of the three extremum solutions in the case of the isothermal
sphere ($\Gamma =0,$ $\protect\beta =2$). Only the least concentrated
solution, which has the highest value of $\Psi _{0}$, is stable. The lower
the value of $\Psi _{0,}$ the more condensed and unstable the solutions
become. The unstable solutions are characterized by a pronounced core-halo
structure.}}{\Qlb{FIG03}}{fig03.eps}{}
In
order to see the difference between density profiles of the isothermal
sphere ( $\Gamma =0$) and the others, we plot the high temperature solutions
(Figure \ref{FIG04}). These are the $\Psi _{0}\rightarrow \infty $
asymptotic solutions where $\beta \rightarrow 0$. If $\Gamma >0$ the
relative volume density profiles curve down, but when $\Gamma <0$ the
profiles curve up, indicating that if we have an $l^{2}$ reservoir, at
higher radius the density profile should change from the homogeneous density
profile.\newpage

In order to show how the $L_{2}$ constraint affects the shape of the density
profiles, in Figure. \ref{FIG04} we plot the relative volume density
profiles for a high temperature. As we can see in the figure, when $\gamma
\neq 0$ the density profiles are no longer uniform and, depending on the
sign of $\gamma $, the density is either increasing $\left( \Gamma <0\right) 
$ or decreasing $\left( \Gamma >0\right) $ while in the standard case $%
\left( \Gamma =0\right) $ in the limit of high temperature, the density is
uniform. We can understand this behaviour if we recognize that in the limit $%
\beta \rightarrow 0$ gravity can be neglected. Consider \ref{3dpointdensity}
when $\gamma \neq 0$. Since the kinetic energy contribution is still
Maxwellian, the probalility of finding a particle with large $l$ is smaller
when $\Gamma >0$ $\left( \gamma >0\right) ,$ which means fewer particles
will occupy larger radii. From the physical point of view, for a relatively
large $L_{2},$ more particles should concentrate at larger radii in order to
maintain the large value, while for relatively small $L_{2},$ fewer
particles should settle at large radii in order to balance the centrifugal
forces. For the case where $\Gamma <0$ the situation is the opposite, and
the density profile should increase with increasing radius. Of course, while
we cannot use this argument for finite temperatures, the origin of the
difference in the density profiles when $\Gamma \neq 0$ is due to this
effect . At finite temperature the tendency persists but the behavior is not
guaranteed.\FRAME{ftFU}{3.9946in}{3.0312in}{0pt}{\Qcb{In order to show the
effect that a nonzero value of $\protect\gamma $ has on the density, we plot
the relative volume mass density profiles of the locally stable
high-temperature asymptotic solutions ($\Psi _{0}\rightarrow \infty $) at
different values of $\Gamma $ (from the top to the bottom, $\Gamma
=-0.5,0,2,5$). While the isothermal sphere has a constant density in this
limit, nevertheless the models with $\Gamma \neq 0$ have different behavior
due to the non-zero $\protect\gamma $.}}{\Qlb{FIG04}}{fig04.eps}{}\newpage

\FRAME{ftbpFU}{3.9946in}{3.0312in}{0pt}{\Qcb{The $\protect\beta $ vs $E$
plot of the extremum solutions when $L_{2}=0.25$ ($E-L_{2}$ ensemble). As
seen, there are no solutions below a critical value of the energy $%
E_{c}\left( L_{2}\right) $, and multiple solutions are present for a certain
range of $E$. Based on the stability investigations, only the upper envelope
of the complete spiraling curve $\Gamma \neq 0$ represents the stable
solutions. Other pieces are unstable (more eigenvalues become negative) as
we go to the center of the spiral. For comparison, see Figure 7.}}{\Qlb{FIG05%
}}{fig05.eps}{}In Figures \ref{FIG05} and \ref{FIG06}%
, the stability properties are presented according to the previously
discussed Poincare's linear series of equilibria with fixed $L_{2}$ and $E$
in the $E-L_{2}$ ensemble. As we can see, the gravothermal catastrophe holds
for the $E-L_{2}$ ensemble as well. In Figure \ref{FIG05}, below a critical
energy there's no extremum solution (here the radius is fixed, not the
energy as in\cite{lynden1}). Also, at the same point, $\dot{\beta}_{a}\left(
E_{c}\right) =\infty $ jumps to $\dot{\beta}_{b}\left( E_{c}\right) =-\infty
.$ ( We follow the spiral in the counter-clockwise direction.) \newpage

\FRAME{ftbpFU}{3.9946in}{3.0519in}{0pt}{\Qcb{The $\protect\gamma $ vs $L_{2}$
plot of the extremum solutions when $E=-0.3.$($E-L_{2}$ ensemble). Above a
critical value of $L_{2},$ there are no extrema. In this case, the
Gravothermal catastrophe is also present according to the stability
investigations, and only the first piece of the spiraling curve which starts
at $L_{2}=0$ and ends at the critical value of $L_{2}$ defined above
represents the stable solutions.}}{\Qlb{FIG06}}{fig06.eps}{}
Thus only the branch labelled $a$ is stable, and the other
branches are increasingly unstable because additional eigenvalues become
negative. In the opposite case, when we fix $E$ (Figure \ref{FIG06}), we can
see that above a critical value of $L_{2}$ there's no extremum solution. The
corresponding functions, $S\left( E\right) $ and $S\left( L_{2}\right) ,$
are plotted in Figure \ref{FIG07}. We see that the entropy is monotonic up
to a critical energy in the stable region of the extremum solutions.\newpage

\FRAME{ftbpFU}{3.9946in}{3.0519in}{0pt}{\Qcb{The entropy vs. energy curve
for the case of $L_{2}=0.25$ ($E-L_{2}$ ensemble). (Also see Figure 5.) The
stable branch of the solutions in Figure 5 can be identified as the highest
entropy curve segment in the $S$ vs $E$ plot, which exactly ends at $E_{c}.$
Note that, this segment is also a monotonic function of $E,$ which means no
phase transition is present in the system.}}{\Qlb{FIG07}}{fig07.eps}{}In the $T-\gamma $ ensemble, the results are presented in Figures 
\ref{FIG08} and \ref{FIG09}. First of all, in the case of fixed $\gamma $,
we have to take a close look at the solutions in the $\left( -E\right)
-\beta $ graph. Only the first branch of the solutions are locally stable up
to a critical value of $\beta $ , say $\beta _{c\text{,}}$ because $\left( -%
\dot{E}\right) _{a}\left( \beta _{c}\right) =\infty $, and there's no
extremum for $\beta >\beta _{c}$. As a comparison, we selected a value of $%
\Gamma $ in Figure \ref{FIG02} corresponding to Figure \ref{FIG08}; as we
wind along the curve to a particular $\beta _{0}$ value, the extremum
solutions are represented by a more concentrated set of density profiles.
Another result of the stability investigations is that in Figure \ref{FIG01}%
, only those solutions are locally stable which are to the right of the
largest maximum of the $\beta \left( \Psi _{0}\right) $ curves. The
corresponding -$F$ can be seen in Figure \ref{FIG10}. In the stable region,
it's a monotonic function, and therefore there's no sign of a phase
transition. The same holds in the $E-L_{2}$ ensemble as can be seen by
inspecting the entropy curves in Figure \ref{FIG07}. The results for fixed $%
\beta $ are shown in Figure \ref{FIG09}. The stable region of extremum
solutions becomes unstable at $\gamma _{c}$ and at $\gamma >\gamma _{c}$
there are no extrema. The free energy behaves similarly to the case of fixed 
$\gamma $ : it's monotonic, and there's no phase transition.\newpage

\FRAME{fpFU}{3.9946in}{3.0519in}{0pt}{\Qcb{The $\left( -E\right) $ vs $%
\protect\beta $ plot of the extremum solutions in the $T-\protect\gamma $
ensemble at $\protect\gamma =-1.$ There are no extrema above the critical $%
\protect\beta _{c}\left( \protect\gamma \right) ,$ and only the first branch
of the extrema are stable. That piece of the spiraling curve which
represents the stable solutions starts at $\protect\beta =2$ and ends at $%
\protect\beta _{c}$ (Gravothermal catastrophe in the $T-\protect\gamma $
ensemble) From Figure 2, we can clearly see that only the least condensed
density profiles represent the stable solutions. For comparison, see Figure
10.}}{\Qlb{FIG08}}{fig08.eps}{}\FRAME{fpFU}{3.9946in}{3.0519in}{0pt}{%
\Qcb{The $\left( -L_{2}\right) $ vs $\protect\gamma $ plot of the extremum
solutions at $\protect\beta =2$ ($T-\protect\gamma $ ensemble). There are no
extrema above the critical value, $\protect\gamma _{c}\left( \protect\beta %
\right) ,$ and only the first branch of the extrema are stable. The curve
segment which represents the stable solutions starts at $\protect\gamma =-1$
and ends at $\protect\gamma _{c}.$ In other words, the Gravothermal
catastrophe is present.}}{\Qlb{FIG09}}{fig09.eps}{}\FRAME{fpFU}{3.9946in}{3.0519in}{0pt}{\Qcb{$\left( -F\right) $ vs $%
\protect\beta $ plot at fixed $\protect\gamma =-1,$ where $F$ is the
thermodynamical potential in the $T-\protect\gamma $ ensemble$.$ From Figure
8, we can clearly identify which piece of the curve represents the stable
solutions, which is the first piece, starting at $\protect\beta =2$ and
ending at $\protect\beta _{c}.$ There is no sign of a phase transition
because in the stable region $-F\left( \protect\beta \right) $ is monotonic.
Note that the stable solutions have the minimal $F.$}}{\Qlb{FIG10}}{fig10.eps%
}{}

\section{Conclusion}

The main purpose of this work is to study the equilibrium thermodynamics of
spherically symmetric self-gravitating systems in the mean field limit. We
investigated both the spherically symmetric point mass system and shell
systems of differing dimension confined in a sphere. These systems are
related to each other both in the form of their Hamiltonian and their
equilibrium states. Furthermore, the \textit{three} dimensional shell system
has the interesting and potentially useful property that, with regard to the
equilibrium density profile, it is equivalent to the spherically symmetric
point mass system. Our description is more general than the standard
treatment of the isothermal sphere \cite{chandra}\ since, in addition to the
energy, we take into account $L_{2}$, the sum of the squares of the
individual angular momentum, which is conserved in the mean-field limit for
spherically symmetric systems. In this new type of microcanonical
description ($E-L_{2}$ ensemble) we evaluated the ''equilibrium''
one-particle probability density function for each type of system by finding
the extrema of the entropy. The resulting pdf's turned out to be similar to
Eddington's anisotropic density function.\cite{Eddington} Therefore the
density profiles obtained here also differ from the isothermal sphere which,
in our formulation, is the special case $\gamma =0.$ Near the system center,
the density profiles are similar to those of the isothermal sphere. However,
as the outer boundary is approached, depending on the value of $\Gamma $,
deviations can become large, increasing or decreasing depending on the sign
of $\Gamma $. The physics behind this behavior is simple: if the system is
spun up corresponding to negative $\Gamma $, the outer density increases.
If, on the other hand, the radial kinetic energy dominates the rotational
energy, $\Gamma \gtrdot 0$ and the outer density decreases.

In addition to the microcanonical ensemble, the Lagrange multiplier $\gamma
, $ which arose from the constraint on $L_{2},$ yielded a new type of
canonical ensemble ($T-\gamma )$ which corresponds to the system being
embedded in a heat bath at temperature $T$ and an $l^{2}$ reservoir at $%
\gamma $. But this analogy is only correct if the system is, at the least,
in a local equilibrium state, and this issue demonstrates the importance of
checking the stability properties of the extremum solutions. The method
known as Poincare's linear series of equilibria proved adequate to analyze
the stability of the extremum solutions in both ensembles with a little
extra effort. Using it, we showed that only certain types of solutions are
locally stable, and others are saddle points. In other words, the
gravothermal catastrophe is also present both in the $E-L_{2}$ and $T-\gamma 
$ ensembles, which means that there is no phase transition in our spherical
mean-field models, although one was expected to be present because of the $%
L_{2}$ constraint.

From their description of the gravothermal catastrophe it is easy to imagine
that Lynden-Bell and Wood had in mind a dynamical process in which mass was
transferred from the halo to a concentrated central core. However, their
approach was confined to a comparison of stationary states. Hints of
collapse have been seen in some N particle simulations.\cite{El-Zant} In
future work we plan to study the complete dynamics of the collapse, both
analytically and using dynamical simulation of N particle and N shell
systems. The shell systems should prove especially useful since they avoid
the complication arising from the formation of tight binaries\cite{binney}.

\section{Acknowledgments}

The authors benefitted from conversations with Igor Prokhorenkov and Michael
Kiessling. They also are grateful for the support of the Research Foundation
and Department of Information Services of Texas Christian University.

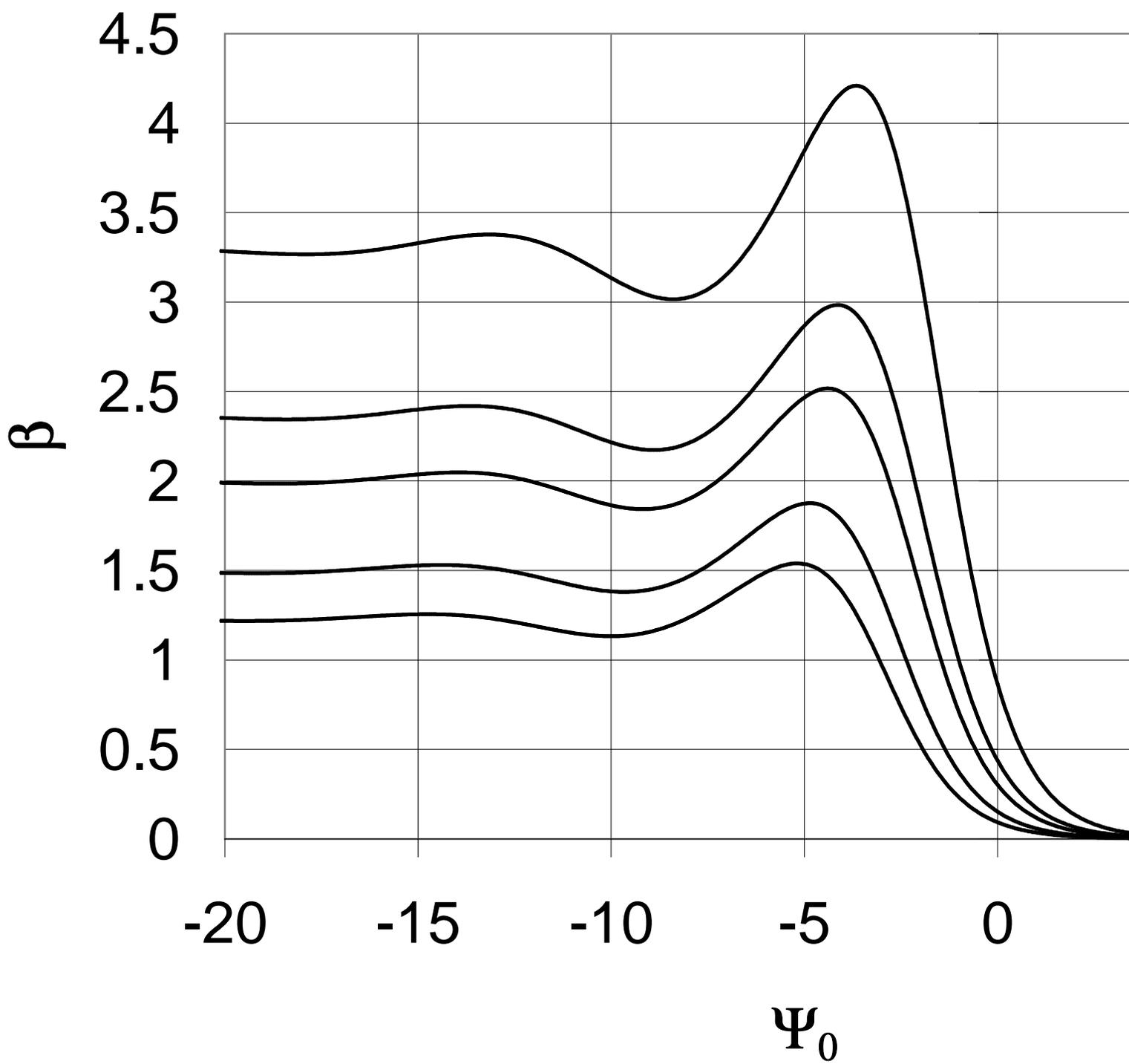
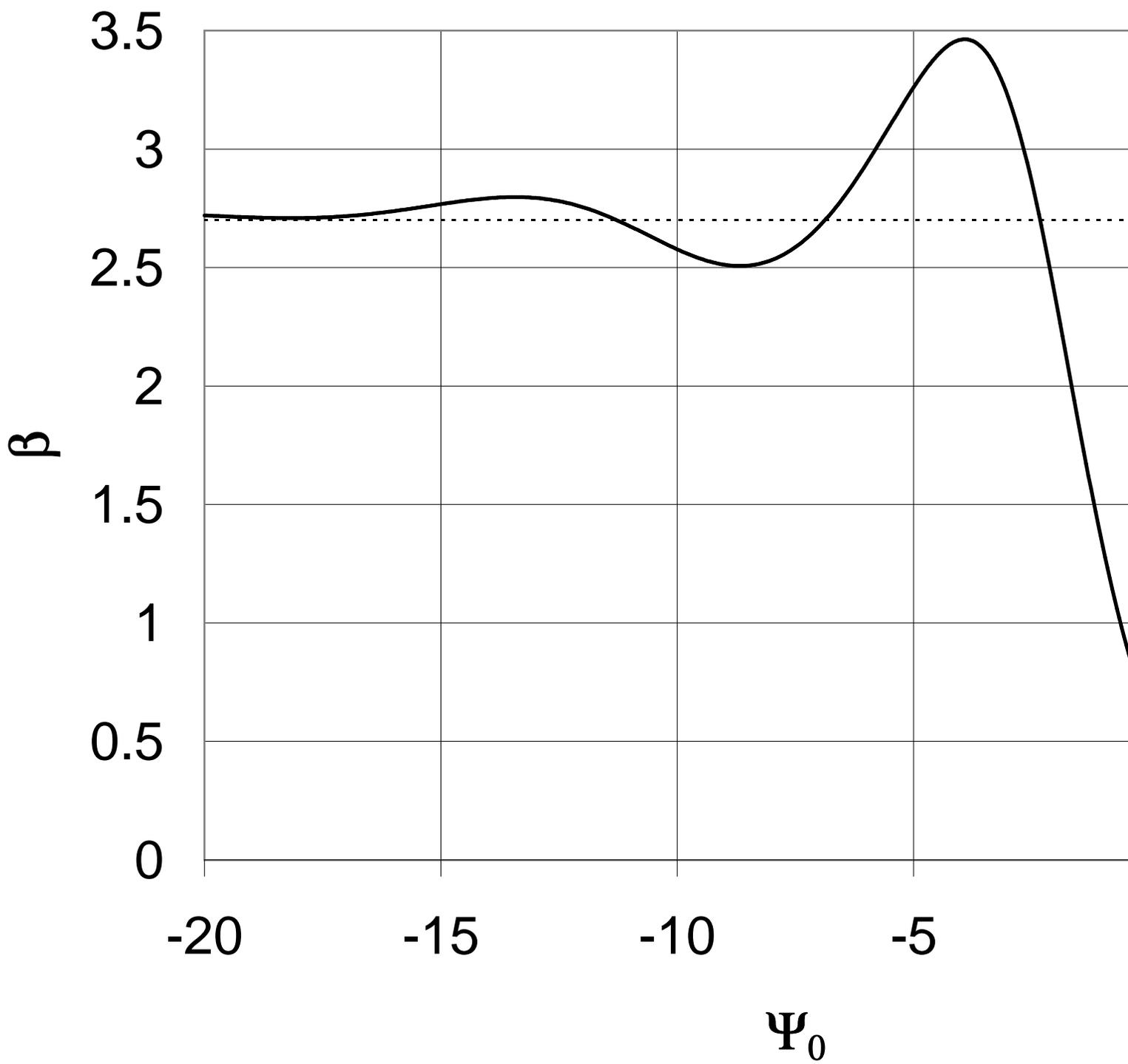
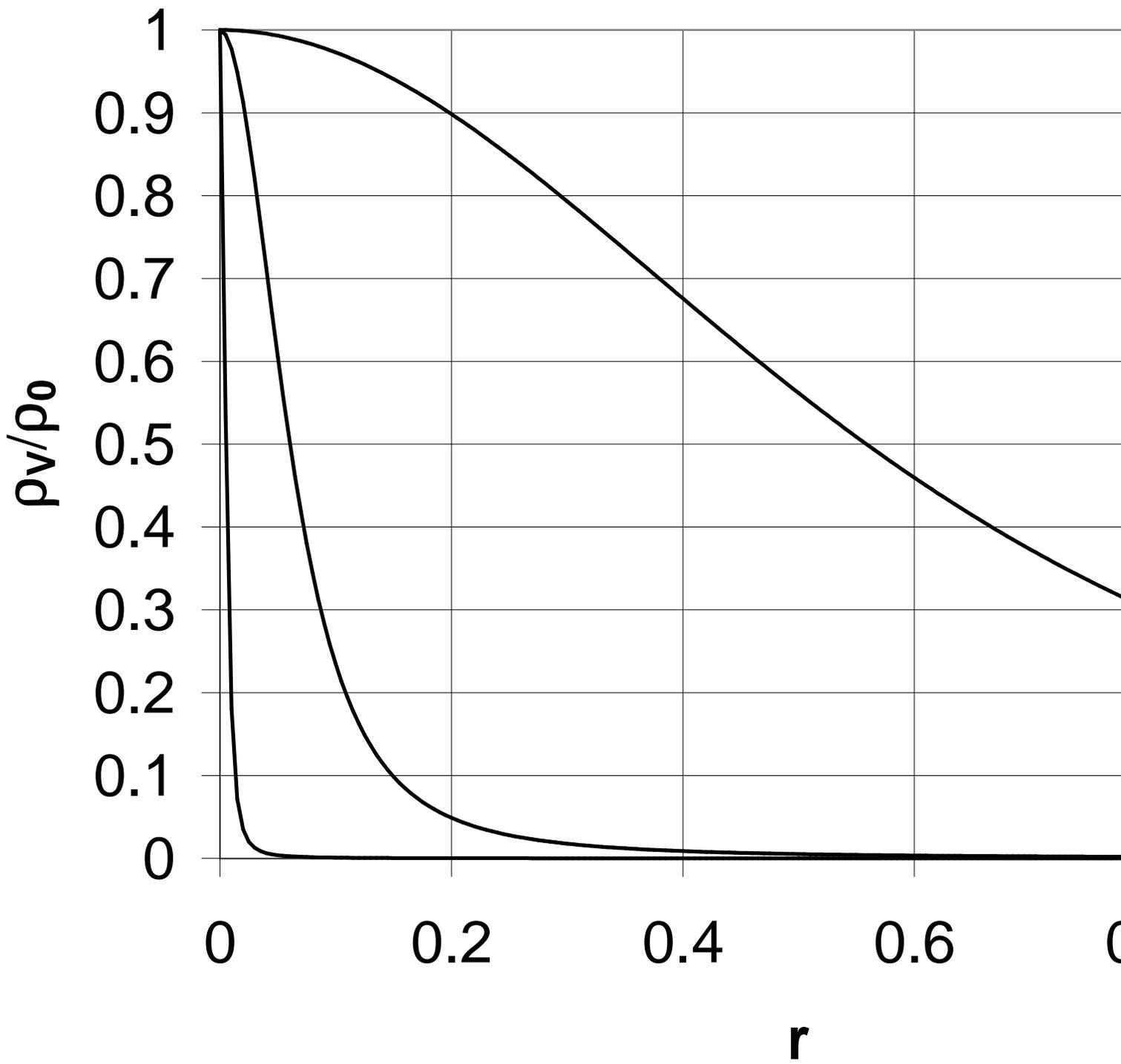
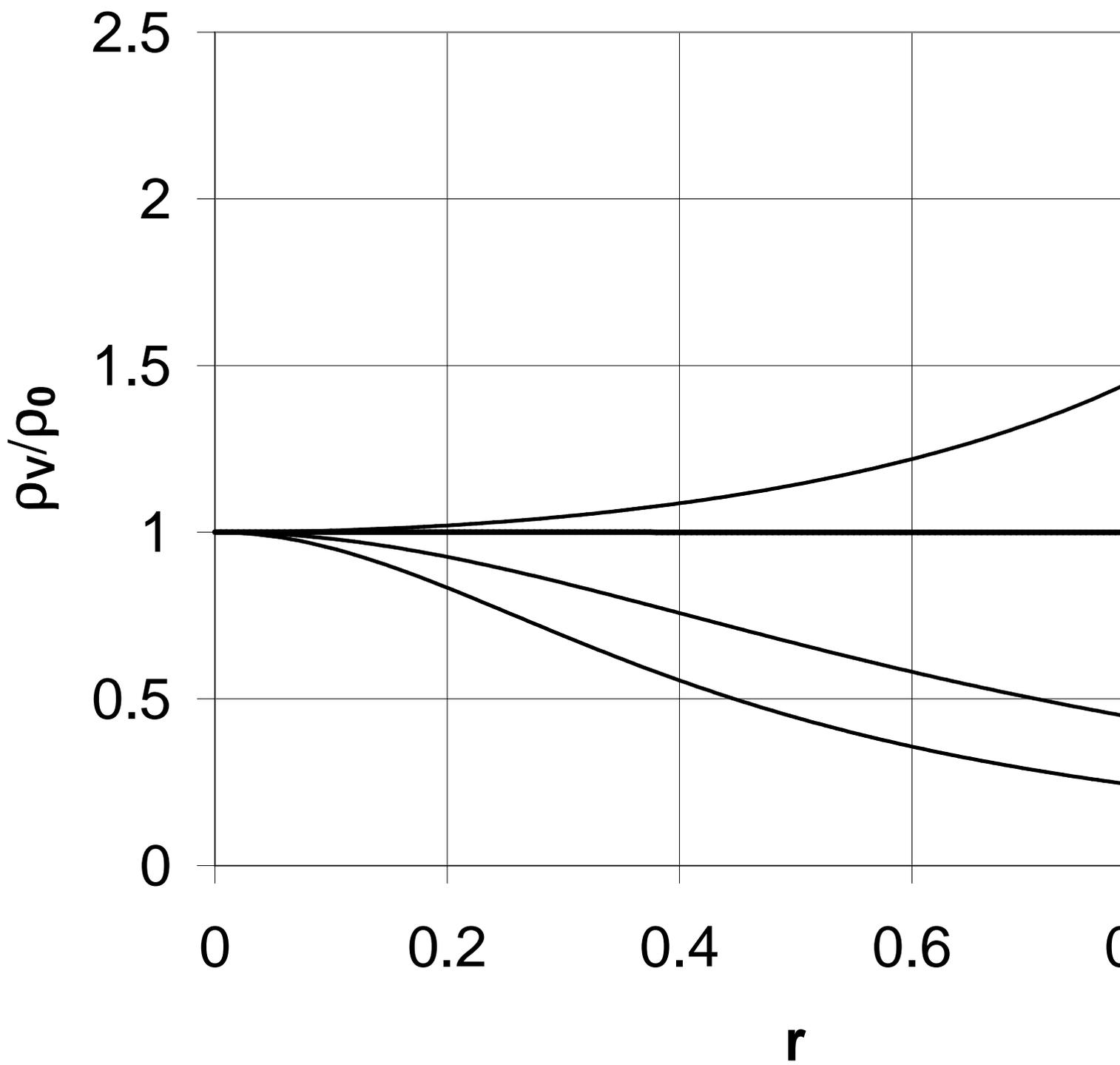
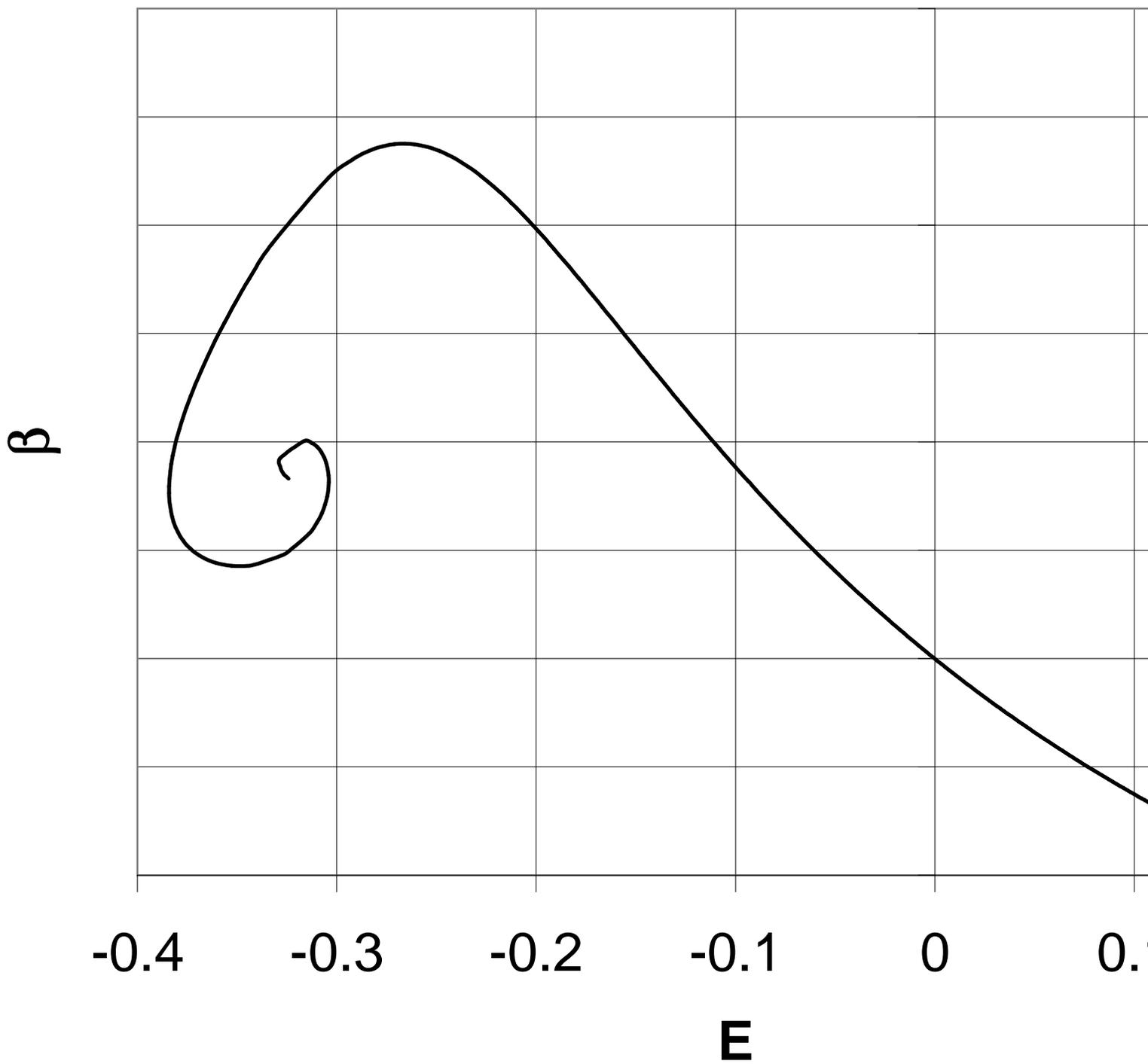
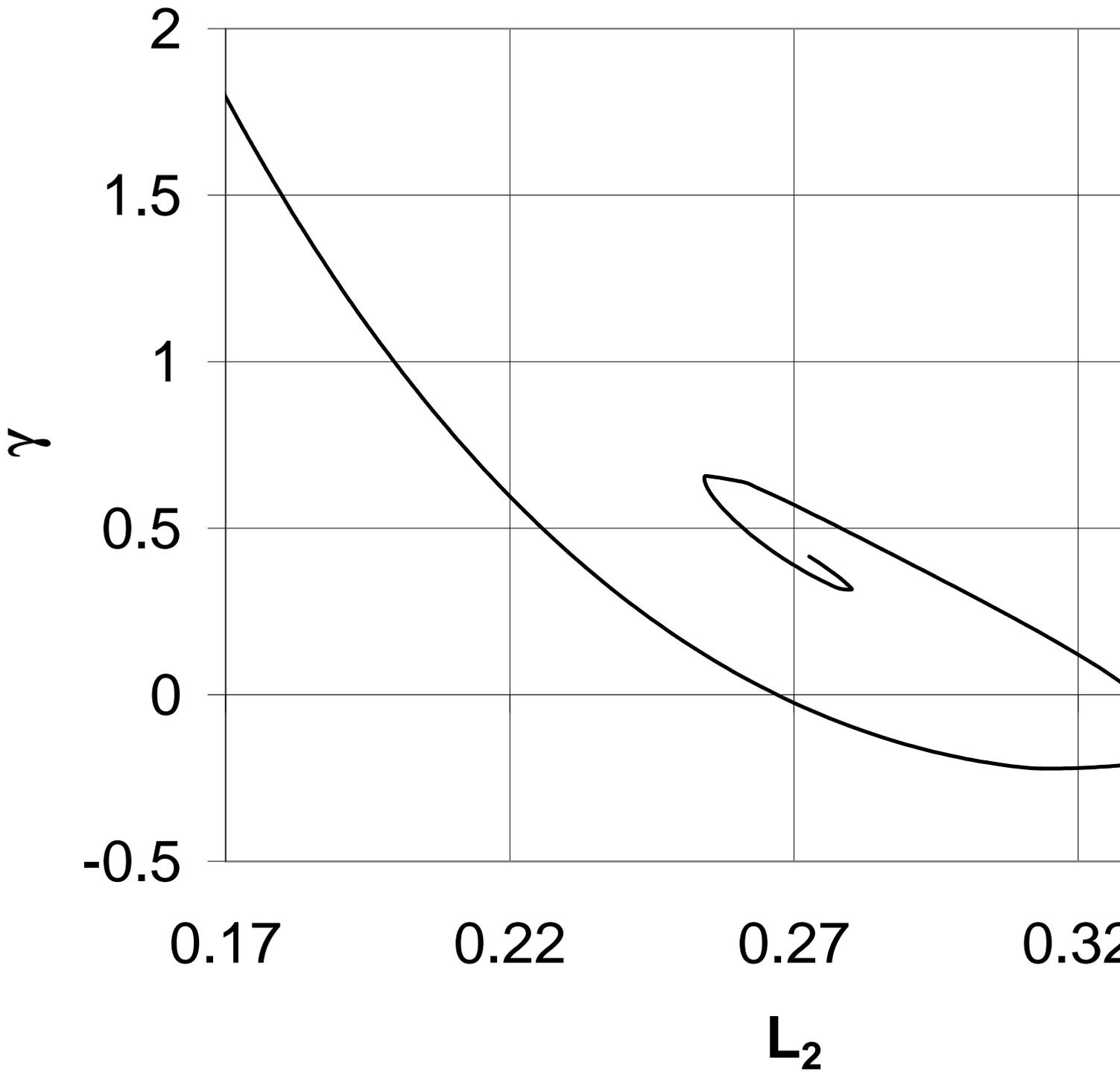
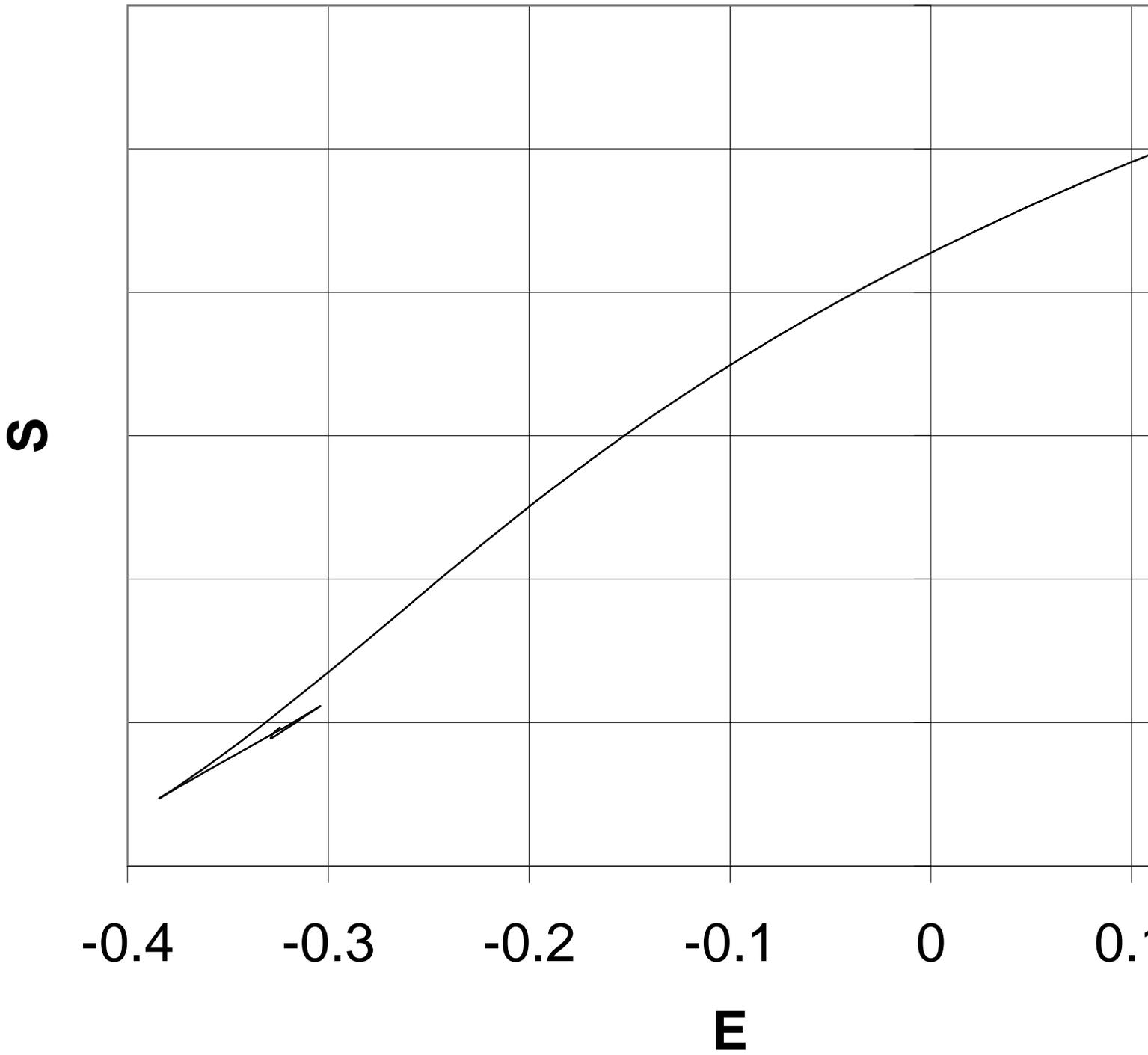
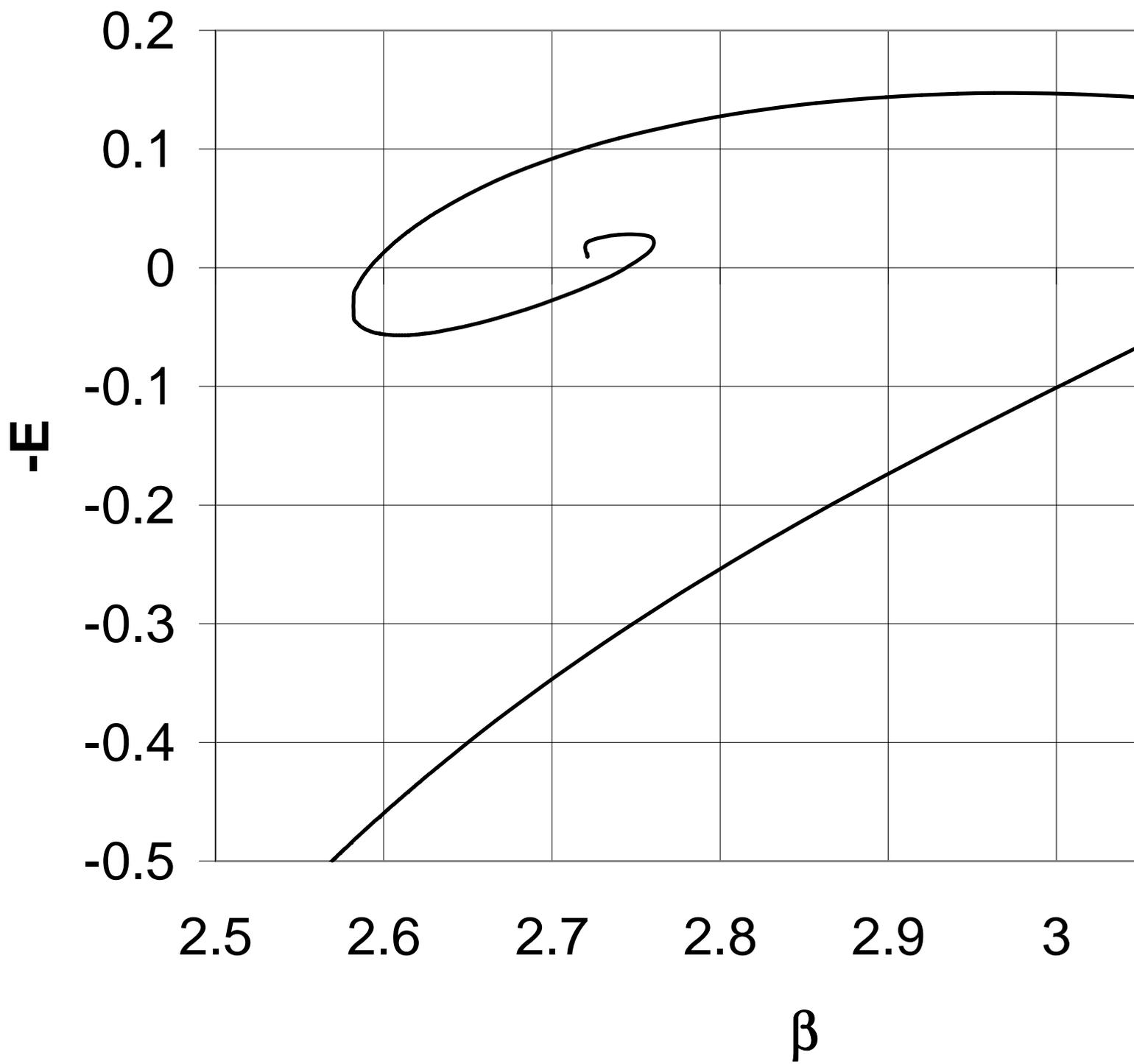
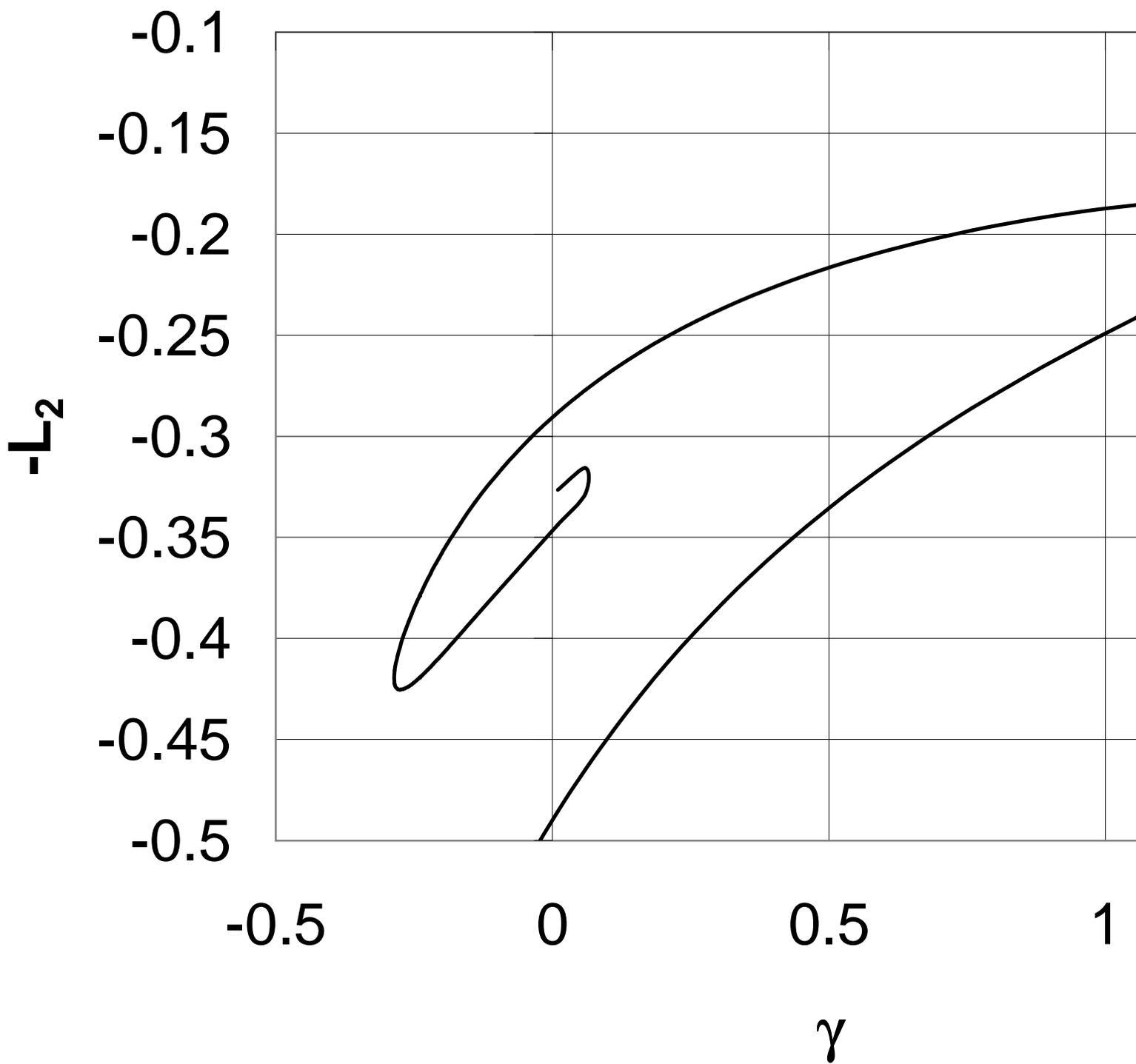
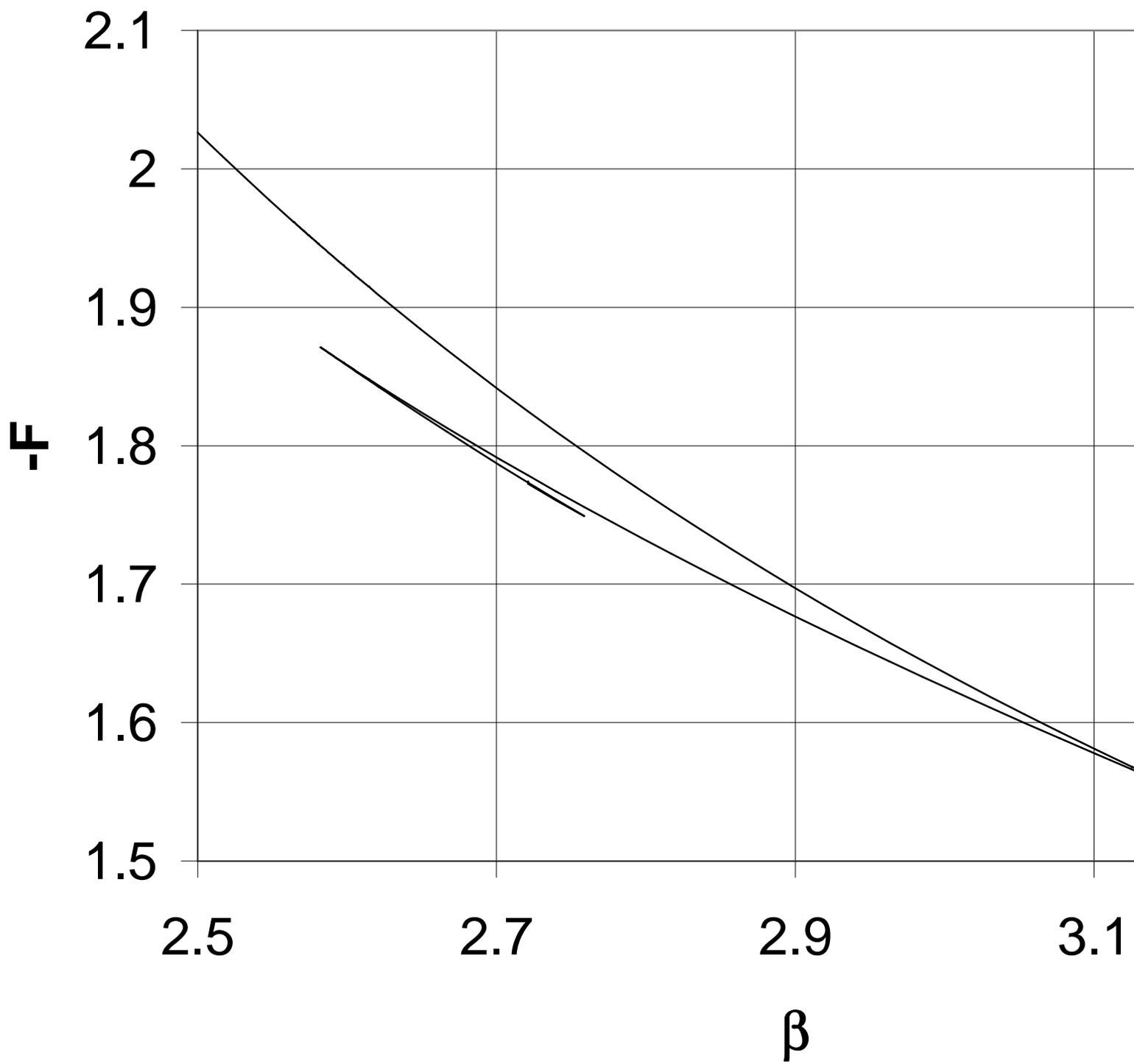
\end{document}